\documentclass[aps,prl,reprint,amsmath,amssymb,superscriptaddress,longbibliography]{revtex4-2}

\usepackage{graphicx}
\usepackage{bm}
\usepackage{dcolumn}
\usepackage{gensymb}
\usepackage[utf8]{inputenc}
\usepackage{float}

\usepackage{url,hyperref,breakurl}

\usepackage{array,mathtools,amssymb,booktabs}
\newcolumntype{C}{>{$}c<{$}}
\AtBeginDocument{
\heavyrulewidth=.08em
\lightrulewidth=.05em
\cmidrulewidth=.03em
\belowrulesep=.65ex
\belowbottomsep=0pt
\aboverulesep=.4ex
\abovetopsep=0pt
\cmidrulesep=\doublerulesep
\cmidrulekern=.5em
\defaultaddspace=.5em
}

\usepackage{hyperref}

\newcommand{\tref}[1]{Table~\ref{#1}}
\newcommand{\eref}[1]{Eq.~(\ref{#1})}

\begin{document}

\title{Precision calculations of the hyperfine constants and isotope and isomer shifts in low-lying states of
$^{229}$Th$^{3+}$}

\author{S. G. Porsev}
\affiliation{Department of Physics and Astronomy, University of Delaware, Newark, Delaware 19716, USA}
\author{M. S. Safronova}
\affiliation{Department of Physics and Astronomy, University of Delaware, Newark, Delaware 19716, USA}

\begin{abstract}
We present high-precision calculations of the hyperfine constants, isotope shift, and isomer shift in low-lying states of Th$^{3+}$ relevant to
spectroscopy of the $^{229}$Th nuclear isomer. The removal energy and hyperfine constants are calculated using a relativistic coupled-cluster method including single, double, and triple excitations, demonstrating the importance of higher-order correlation effects. The isotope shift of
the $5f_{5/2}-7s$ transition is evaluated using both the CI+MBPT and CI+all-order methods, providing an estimate of the theoretical uncertainty.
We find that the isotope shift is dominated by the field shift, with the mass shift contributing negligibly. Combining the calculated field-shift
coefficient with the measured difference in the nuclear mean-square charge radii of the ground and isomeric states, we predict the $7s$ isomer shift
to be $\Delta E^{m,g}=0.9(1)$ GHz. These results provide benchmark atomic data for Th$^{3+}$ and support ongoing precision spectroscopy
of $^{229}$Th and the development of a nuclear clock.
\end{abstract}

\date{\today}

\maketitle
\section{Introduction}
The low-energy nuclear isomeric state in $^{229}$Th has attracted considerable interest due to its potential application in a nuclear clock and provides a unique bridge between atomic and nuclear physics~\cite{PeiTam03,ThiOkhGlo18}. With an excitation energy in the vacuum-ultraviolet range, recently determined to be 8.3557 eV~\cite{TieOkhZha24,ElwSchJee24}, the $^{229{\rm m}}$Th isomer is the only known nuclear state accessible to direct laser excitation. This property underlies proposals for a nuclear frequency standard with potentially enhanced stability and reduced sensitivity to external perturbations compared with atomic clocks~\cite{PeiTam03,CamRadKuz11,PeiSchSaf21}.

Recent experimental progress has enabled direct studies of the isomeric transition. The radiative decay of $^{229{\rm m}}$Th has been observed~\cite{KraMoeAtn23}, followed by laser excitation of the nuclear transition~\cite{TieOkhZha24,ElwSchJee24}, and a frequency measurement referenced to the $^{87}$Sr optical clock~\cite{ZhaOoiHig24}. The first realization of a closed feedback loop and actual clock operation based on the $^{229}$Th isomer transition was reported in  Refs.~\cite{HuaYanXia26,TosRieMor26}.

These results provide quantitative constraints on the nuclear transition energy and motivate increasingly accurate theoretical descriptions of electronic structure effects relevant for interpretation of spectroscopic data.

Among thorium ions, Th$^{3+}$ is of particular interest due to its single-valence-electron structure, which enables accurate treatment using relativistic many-body methods. This system has therefore been widely used as a benchmark for testing high-precision atomic-structure calculations and for extracting nuclear properties from spectroscopic observables~\cite{SafSafRad13,PorSafKoz21}. Reliable determination of energies, hyperfine-structure (HFS) constants, isotope shifts, and isomer shifts is required for quantitative analysis of present and future measurements.

Relativistic coupled-cluster and configuration-interaction methods have been applied extensively to Th$^{3+}$ and related heavy atomic systems. Calculations of energies, transition amplitudes, polarizabilities, and HFS constants have demonstrated the essential role of nonlinear correlation contributions and triple excitations~\cite{SafSafRad13,PorSaf21,PorSafKoz21}. Similar all-order many-body techniques have been used in studies of heavy systems relevant to precision tests of fundamental physics, including Cf ions proposed for optical-clock applications and investigations of possible variation of the fine-structure constant~\cite{PorSafSaf20,PorSaf26}. Together, these studies demonstrate the applicability of the same theoretical framework across different heavy atomic systems. In parallel, measurements of isotope shifts and hyperfine structure in thorium ions have provided information on nuclear moments and changes in nuclear charge radii~\cite{CamRadKuz11,SafPorKoz18,ZitTieDul25,YamShiHab24}, including the difference in mean-square charge radii between the ground and isomeric nuclear states of $^{229}$Th~\cite{SafPorKoz18}.

In the present work we perform high-precision calculations of properties of Th$^{3+}$ relevant to spectroscopic studies of the $^{229}$Th isomer. We calculate the removal energy and magnetic-dipole HFS constant of the $7s$ state, as well as the magnetic-octupole HFS constant of the ground $5f_{5/2}$ state, using a relativistic coupled-cluster method that includes single, double, and triple excitations~\cite{PorDer06,PorSafKoz21,PorSafSaf20,PorSaf26}. This approach allows systematic inclusion of higher-order correlation effects and enables quantitative assessment of their contribution.

Isotope shift of the $5f_{5/2}-7s$ transition is evaluated using both CI+MBPT and CI+all-order methods~\cite{DzuFlaKoz96,SafKozJoh09}. Comparison of the two approaches provides an estimate of the higher-order correlation effects and yields the field-shift coefficient with improved reliability. Combining this result with the experimental difference in nuclear mean-square charge radii between the ground and isomeric states~\cite{SafPorKoz18}, we determine the isomer shift of the $7s$ level in $^{229}$Th$^{3+}$.

The results provide benchmark data for atomic structure calculations in Th$^{3+}$ and contribute to the theoretical support required for precision spectroscopy of the $^{229}$Th nuclear isomer and related applications in nuclear-frequency metrology.
\section{Method of calculation}
We calculate the energy and the HFS constant $A$ for the $7s$ state, as well as the HFS constant $C$ for the ground $5f_{5/2}$ state, using a version of the high-precision relativistic coupled-cluster method developed in Ref.~\cite{PorDer06} and extended to include core triple excitations~\cite{PorSafKoz21}. The quadratic nonlinear (NL) terms are included in the equations for single and double excitations, but omitted from the equations for triples. Cubic and higher-order terms were also neglected, as they are expected to be small.

We treated Th$^{3+}$ as a monovalent ion. The initial Dirac-Hartree-Fock (DHF) self-consistency procedure included the Breit interaction and is carried out for the closed-shell core. The coupled-cluster equations were solved in a basis set of single-particle orbitals. In the equations for singles, doubles, and valence triples, the sums over excited states included 35 basis orbitals with orbital quantum number $l \leq 6$.

Because of the high computational cost associated with the iterative solution of the core-triple equations, several restrictions were imposed
because unrestricted iterative treatment of core triples remains computationally prohibitive for the present basis. Core excitations were allowed only from the $[4d\text{--}6p]$ shells and the maximum orbital quantum number of the excited orbitals was limited to $l_{\max}=5$.
\section{Energies.}
Numerical results for the removal energy of the valence $7s$ state, obtained at different stages of the calculation, are presented in Table~\ref{Tab:E}.
The lowest-order DHF contribution to the energies (including the Breit interaction) is labeled ``BDHF''. At the next step, we carried out calculations
in the linearized coupled-cluster single-double (LCCSD) approximation~\cite{BluJohLiu89,BluJohSap91}. We then successively included the nonlinear (NL) terms~\cite{EliKalIsh96,PalSafJoh07}, valence triples~\cite{PorDer06,SahDasCha07,PorSaf21},
and core triples~\cite{PorSafKoz21,PorSaf26}, designating the corresponding calculations as CCSD, CCSDvT, and CCSDT, respectively. Thus, each subsequent calculation includes all
terms taken into account at the previous stage together with the additional terms specific to the given approximation. In this way, the most complete calculation was performed in the CCSDT approximation.

We additionally evaluate the complementary basis-extrapolation and quantum-electrodynamic (QED) corrections, $\Delta E_{\rm extrap}$ and $\Delta E_{\rm QED}$.
The former accounts for the contribution of higher partial waves with $l > 6$. Following the empirical rule established for Ag-like ions in Ref.~\cite{SafDzuFla14PRA1} and previously employed in Refs.~\cite{PorSaf21,PorSafKoz21}, we estimated this contribution as the difference between two calculations performed with $l_{\rm max}=6$ and $l_{\rm max}=5$. The QED corrections were evaluated following the approach of Ref.~\cite{TupKozSaf16}.
We find that $\Delta E_{\rm QED}$ exceeds $\Delta E_{\rm extrap}$ by approximately a factor of three in absolute value. The experimental value of the removal energy for the $7s$ state was taken from Refs.~\cite{RalKraRea11,ThIV}.

To demonstrate the systematic improvement of the results as additional coupled-cluster terms are included, we present in the lower panel
of Table~\ref{Tab:E} the differences between the theoretical and experimental values obtained at each stage of the calculation.
Comparing $\Delta_{\rm LCCSD}$ with $\Delta_{\rm CCSDT}$, one sees that the discrepancy between theory and experiment decreases
from 1444 to 568 cm$^{-1}$ by a factor of 2.5 for the $7s$ state when the NL terms and triples are included.
\begin{table}[h]
\caption{The removal energies of the $7s$ state for Th$^{3+}$ (in cm$^{-1}$) in
different approximations, discussed in the text, are presented. The theoretical total and experimental results
are given in the rows $E_{\rm total}$ and $E_{\rm expt}$.
The difference between the total and experimental values is presented (in \%) in the row labeled ``Diff. (\%)''.}
\label{Tab:E}
\begin{ruledtabular}
\begin{tabular}{lr}
\smallskip
                                        &   $7s$   \\
\hline \\[-0.7pc]
$E_{\rm BDHF}$                          &  200183  \\[0.2pc]
$E_{\rm LCCSD}$                         &  209504  \\[0.2pc]
$E_{\rm CCSD}$                          &  209150  \\[0.2pc]
$E_{\rm CCSDvT}$                        &  208672  \\[0.2pc]
$E_{\rm CCSDT}$                         &  208628  \\[0.2pc]
$\Delta E_{\rm QED}$                    &    -187  \\[0.4pc]
$\Delta E_{\rm extrap}$                 &      59  \\[0.4pc]

$E_{\rm total}$                         &  208500  \\[0.3pc]
$E_{\rm expt}$~\cite{RalKraRea11,ThIV}  &  207932  \\[0.1pc]
Diff. (\%)                              &    0.27  \\
\hline \\[-0.6pc]
${\Delta_{\rm LCCSD}}^{\rm a}$          &    1444  \\[0.2pc]
${\Delta_{\rm CCSD}}$                   &    1090  \\[0.2pc]
${\Delta_{\rm CCSDvT}}$                 &     612  \\[0.2pc]
${\Delta_{\rm CCSDT}}$                  &     568
\end{tabular}
\end{ruledtabular}
\begin{flushleft}
$^{\rm a}\!\Delta_X \equiv E_X + \Delta E_{\rm extrap} + \Delta E_{\rm QED} - E_{\rm expt}$.
\end{flushleft}
\end{table}
\section{Hyperfine structure constants}
The magnetic-dipole and electric-quadrupole HFS constants, $A$ and $B$, for the $5f_{5/2,7/2}$ and $6d_{3/2,5/2}$ states of $^{229}$Th$^{3+}$ were
previously calculated in Ref.~\cite{PorSafKoz21}. In the present work, we calculate the magnetic-dipole HFS constant $A$ for the $7s$ state and the
magnetic-octupole HFS constant $C$ for the $5f_{5/2}$ state.

The results for the reduced magnetic-dipole HFS constant $A_t \equiv A/g$ (where $g=(\mu/\mu_N)/I$ is the nuclear $g$ factor, $\mu$ is the nuclear
magnetic moment, $\mu_N$ is the nuclear magneton, and $I$ is the nuclear spin, with $I=5/2$) and the reduced magnetic-octupole HFS constant
$C_t \equiv C/\Omega$ (where $\Omega$ is the magnetic octupole moment) are presented in \tref{Tab:hfs}.
\begin{table}[h]
\caption{Contributions to the HFS constants $A_t$ (in MHz) and $C_t$ (in KHz/(b $\times\, \mu_N$), where b denotes barn). The individual contributions are explained in the text. Uncertainties are given in parentheses.}
\label{Tab:hfs}
\begin{ruledtabular}
\begin{tabular}{lrcc}
                                 & $C_t(5f_{5/2})$  & $A_t(7s)$    \\
\hline \\[-0.6pc]
BDHF                             &      -0.69       &   33280       \\[0.4pc]

$\Delta$(SD)                     &       1.33       &    7929       \\[0.1pc]
LCCSD                            &       0.64       &   41209       \\[0.4pc]

$\Delta$(NL)                     &      -0.20       &    -246       \\[0.1pc]
$\Delta$(vT)                     &       0.39       &    -238       \\[0.1pc]
$\Delta$(cT)                     &      -0.04       &     108       \\[0.1pc]
CCSDT                            &       0.79       &   40833       \\[0.3pc]

QED correction                   &                  &    -293       \\[0.3pc]
Basis extrapolation              &      -0.02       &      23       \\[0.3pc]

Total                            &       0.77       &   40563       \\[0.2pc]
Final                            &      $0.77(13)$  &  $40563(364)$ \\[0.2pc]
Ref.~\cite{SafSafRad13}          &                  &   40320       \\[0.2pc]
Ref.~\cite{LiQiaTan21}           &$0.38(22)^{\rm a}$&
\end{tabular}
\end{ruledtabular}
\begin{flushleft}
$^{\rm a}$Owing to the different sign conventions adopted in the literature for the nuclear magnetic octupole moment $\Omega$, the $C_t$ value from Ref.~\cite{LiQiaTan21} is taken as an absolute value for comparison.
\end{flushleft}
\end{table}

The LCCSD and BDHF values, together with their difference, $\Delta$(SD), are given in the upper panel of the table.
Rows 4--6 list the corrections due to the nonlinear terms, $\Delta$(NL), and the valence and core triples, $\Delta$(vT) and $\Delta$(cT), respectively.
The CCSDT values, obtained as the sum of the LCCSD values and the NL, vT, and cT corrections, are presented in the row labeled ``CCSDT''.
As shown, the $\Delta$(cT) correction for the $7s$ state is smaller but comparable in magnitude to the $\Delta$(vT) and $\Delta$(NL) corrections.
The QED and basis-extrapolation corrections are given in the rows labeled ``QED correction'' and ``Basis extrapolation'', respectively.
The values in the row labeled ``Total'' are obtained as the sum of the CCSDT value, the QED and  basis-extrapolation corrections. In the row ``Final'' we present the final values with their uncertainties.

Based on a comparison of theoretical and experimental HFS constants for a number of univalent elements, the authors of Ref.~\cite{SafSafRad13}
proposed a method for estimating the uncertainties of these constants. According to this approach, the uncertainty of the $A$ calculations is
expected to be on the order of 3--6\% of the total correlation correction (defined as the difference between the final and BDHF values),
provided that this correction does not exceed 50\%. Following this prescription, we estimate the uncertainty of $A_t(7s)$ as 5\% of its total
correlation correction. This estimate is in good agreement with the uncertainty analysis presented in Ref.~\cite{SafSafRad13}.
We note that all values of $A_t(7s)$ presented in \tref{Tab:hfs} were obtained assuming a uniformly magnetized spherical nucleus.

 As seen from~\tref{Tab:hfs}, the BDHF value of $C_t(5f_{5/2})$ has the opposite sign to the
 LCCSD result. Consequently, the uncertainty-estimation procedure adopted for $A_t(7s)$ is not appropriate in this case. Instead, we estimate
 the uncertainty of $C_t(5f_{5/2})$ from the difference between the CCSDT and LCCSD values. We also note that the QED corrections are negligibly small and lie well within the estimated uncertainty. Our final result is in reasonable agreement with that reported in Ref.~\cite{LiQiaTan21}. We attribute the difference primarily to the omission of triple excitations in the calculations of Ref.~\cite{LiQiaTan21} .
\section{Isotope shift}
We calculate the isotope shift (IS) of the $5f_{5/2}\, -\, 7s$ transition frequency in Th$^{3+}$.
In this calculation we employ two complementary approaches that combine configuration interaction with (i) many-body perturbation theory (MBPT)~\cite{DzuFlaKoz96} and (ii) the linearized coupled-cluster (all-order) method~\cite{SafKozJoh09}.
These will be referred to as the CI+MBPT and CI+all-order methods, respectively.
We note that in this case the CI is trivial because there is only one valence electron.

The isotope shift can be expressed as the sum of the field-shift (FS) and mass-shift (MS) contributions,
\begin{eqnarray}
\Delta \nu^{A'A} &=& \Delta \nu^{A'A}_{\rm FS} + \Delta \nu^{A'A}_{\rm MS} \nonumber \\
                 &=& K_{\rm FS}\, \delta \langle r^{2} \rangle^{A'A} + K_{\rm MS} \left( \frac{1}{A'} - \frac{1}{A} \right)
\label{Del_nu}
\end{eqnarray}
where $K_{\rm FS}$ and $K_{\rm MS}$ are the field- and mass-shift coefficients, respectively, $A'$ and $A$ are the mass numbers of the isotopes, and $\langle r^{2} \rangle$ denotes the mean-square nuclear charge radius. The difference in mean-square radii is defined as
\begin{equation}
\delta \langle r^{2} \rangle^{A'A} = \langle r^{2} \rangle^{A'} - \langle r^{2} \rangle^{A}.
\end{equation}

For Th$^{3+}$, the field shift is approximately two orders of magnitude larger in absolute value than the mass shift. Consequently, the total isotope shift is almost entirely determined by the
field-shift contribution. The FS is evaluated using the \textit{finite-field} method, in which a perturbation proportional to the field-shift operator $H_\mathrm{FS}$ is added to the Hamiltonian:
\[
H \rightarrow H_\lambda = H + \lambda H_\mathrm{FS}.
\]

The scaling parameter $\lambda$ is chosen to be sufficiently large for the field-shift effect to exceed numerical noise but small enough to ensure that the resulting energy shift remains linear in $\lambda$. In the present work, we use $\lambda = \pm 0.01$.

The eigenvalues $E$ are obtained by direct diagonalization of $H_\lambda$. Assuming the nucleus to be a uniformly charged ball of radius $R$, the field-shift coefficient $K_{\mathrm{FS}}$ is determined as~\cite{KorKoz07}
\begin{equation}
K_{\mathrm{FS}} = \frac{5}{6R^2}\, \frac{\partial E}{\partial \lambda}.
\label{der}
\end{equation}
The conversion factor between atomic and SI units for $K_{\mathrm{FS}}$ is $1~\mathrm{a.u.} \approx 2.3497 \times 10^{-3}~\mathrm{GHz/fm^2}$.

The mass shift consists of the normal mass shift and the specific mass shift (SMS). The latter is calculated by modifying the Hamiltonian according to $H \rightarrow H_\lambda= H + \lambda H_\mathrm{SMS}$. The SMS coefficient is then obtained as the corresponding derivative $\partial E/\partial\lambda$.
\begin{table*}[htp]
\caption{\label{tab1} Isotope field shift and specific mass shifts calculations.}
\begin{ruledtabular}
\begin{tabular}{lccccccccc}
\multicolumn{1}{c}{} & \multicolumn{1}{c}{} & \multicolumn{1}{c}{} & \multicolumn{4}{c}{Field shift}
& \multicolumn{1}{c}{$\delta \nu_{\rm{FS}}^{232,229}$} & \multicolumn{1}{c}{$\delta \nu_{\rm{SMS}}^{232,229}$}
& \multicolumn{1}{c}{$\delta \nu^{232,229}$} \\
\multicolumn{1}{c}{} & \multicolumn{1}{c}{Energy~\cite{ThIV}} & \multicolumn{1}{c}{}
& \multicolumn{1}{c}{$\frac{\partial E}{\partial \lambda}$(CI+MBPT)} &\multicolumn{1}{c}{$\frac{\partial E}{\partial \lambda}$(CI+All)}
& \multicolumn{1}{c}{Diff.} & \multicolumn{1}{c}{$K_{\rm{FS}}$(CI+All)} &\multicolumn{2}{c}{} \\
\multicolumn{1}{c}{} & \multicolumn{1}{c}{cm$^{-1}$} & \multicolumn{1}{c}{Term} & \multicolumn{1}{c}{a.u.} & \multicolumn{1}{c}{a.u.} &
\multicolumn{1}{c}{\%} & \multicolumn{1}{c}{GHz/fm$^2$}  & \multicolumn{1}{c}{GHz} & \multicolumn{1}{c}{GHz} & \multicolumn{1}{c}{GHz} \\
\hline \\[-0.6pc]
  State     &   0   & $5f_{5/2}$        &  -0.000527  & -0.000507 & -3.8\%  &  -50.4  &        &       &            \\[0.1pc]
            &  4325 & $5f_{7/2}$        &  -0.000509  & -0.000484 & -5.0\%  &  -48.1  &        &       &            \\[0.1pc]
            &  9193 & $6d_{3/2}$        &  -0.000206  & -0.000219 &  5.7\%  &  -21.7  &        &       &            \\[0.1pc]
            & 14486 & $6d_{5/2}$        &  -0.000191  & -0.000191 & -0.2\%  &  -18.9  &        &       &            \\[0.1pc]
            & 23131 & $7s$              &   0.000915  &  0.000873 & -4.9\%  &   86.7  &        &       &            \\[0.1pc]
Transition  &       & $5f_{5/2}\, - 7s$ &  -0.001442  & -0.001380 & -4.5\%  & -137.1  & -41.0  &  0.2  & -40.8(2.5)
\end{tabular}
\end{ruledtabular}
\end{table*}

The results for the FS ${\partial E}/{\partial\lambda}$ derivatives and coefficients are summarized in Table~\ref{tab1}. We use two different methods for the calculations of the field shift: combination of the CI with many-body perturbation theory (CI+MBPT) \cite{DzuFlaKoz96} and more accurate CI+all-order method \cite{SafKozJoh09}. These approaches allow one to incorporate core excitations in the CI method by constructing an effective Hamiltonian $H_{\mathrm{eff}}$ using either second-order MBPT or linearized coupled-cluster methods, respectively. The CI+all-order method includes third- and higher-order corrections to the effective Hamiltonian. The use of two independent methods allows us to establish the effect of the higher orders and to estimate the accuracy of the final results. The corresponding results for the derivatives are given in the columns labeled ``$\frac{\partial E}{\partial \lambda}$(CI+MBPT)'' and ``$\frac{\partial E}{\partial \lambda}$(CI+All)''. The difference between these results provides an  estimate of the uncertainty of our calculation, listed in the ``Diff.'' column in \%.

To evaluate the FS for the $5f_{5/2}\, - 7s$ transition, we use $\delta \langle r^{2} \rangle^{232,229} = 0.299(15)\,\mathrm{fm^2}$ from Ref.~\cite{SafPorKoz18} and the CI+all-order values of the field-shift coefficients. Considering the differences between the CI+MBPT and CI+all-order results as the uncertainty of our calculation and taking into account the uncertainty of
$\delta\langle r^2\rangle^{232,229}$, we estimate the uncertainty of our calculated isotope shift at the level of 6\%.

As seen from \tref{tab1}, the SMS is more than two orders of magnitude smaller than the FS in absolute value. Given its small contribution, a high-precision calculation of the SMS coefficient is not required, and we calculate it within the CI+MBPT approximation. According to our estimate, the normal mass shift is an order of magnitude smaller than the SMS and is therefore completely negligible.
\section{Isomer shift for the $7s$ state}
Using the coefficient $K_{\rm FS}(7s) = 87(4)\,{\rm GHz/fm^2}$, calculated in the previous section, we can find the isomer shift $\Delta E^{m,g}$ for the $7s$ state, as
\begin{equation}
\Delta E^{m,g} = K_{\rm FS}\, \delta \langle r^{2} \rangle^{m,g},
\label{Del_E}
\end{equation}
where $m$ and $g$ designate the isomeric and ground states of the nucleus of $^{229}$Th$^{3+}$. The uncertainty of $K_{\rm FS}(7s)$ was determined as the difference between the CI+all-order and CI+MBPT values.

The value of $\delta \langle r^{2} \rangle^{m,g}$ was found in Refs.~\cite{SafPorKoz18} and \cite{YamShiHab24} to be 0.0105(13) and 0.0097(26)
 fm$^2$, respectively. Using the former value in ~\eref{Del_E}, we obtain $$\Delta E^{m,g} = 0.9(1)\,{\rm GHz}.$$
\section{Conclusion}
We performed high-precision calculations of several atomic properties of Th$^{3+}$ relevant to spectroscopic studies of the low-energy nuclear isomer
in $^{229}$Th. The removal energy and magnetic-dipole HFS constant of the $7s$ state, as well as the magnetic-octupole HFS constant of the ground $5f_{5/2}$ state, were evaluated using a relativistic coupled-cluster method that includes single, double, and triple excitations.
The calculations demonstrate the importance of nonlinear terms and triple excitations for achieving high accuracy. For the $7s$ removal energy, inclusion
of these higher-order correlation effects reduces the discrepancy with experiment by a factor of approximately 2.5 compared with the linearized
coupled-cluster result.

We obtained the reduced magnetic-dipole hyperfine constant $A_t(7s)=40563(364)$ MHz and the reduced magnetic-octupole hyperfine constant
$C_t(5f_{5/2})=0.77(13)$ kHz/(b $\times \mu_N$). The uncertainty of the $A_t(7s)$ value was estimated using established correlation-based scaling arguments
and is expected to be at the level of 1\%.

The isotope shift of the $5f_{5/2}-7s$ transition was investigated using both the CI+MBPT and CI+all-order methods. Comparison of the two approaches
allowed us to assess the contribution of higher-order correlation corrections and estimate the theoretical uncertainty of the field-shift coefficient.

Combining the calculated field-shift coefficient with the experimentally determined change in the nuclear rms charge radius between the ground
and isomeric states of $^{229}$Th, we determined the isomer shift of the $7s$ level to be $\Delta E^{m,g}=0.9(1)$ GHz.
The calculated $7s$ isomer shift provides a direct theoretical prediction for the frequency displacement associated with the nuclear isomer and may assist in the identification and interpretation of spectroscopic signals from $^{229m}$Th$^{3+}$.

The results presented here provide benchmark data for atomic-structure calculations in Th$^{3+}$ and contribute to the theoretical support required
for ongoing and future precision spectroscopy of the $^{229}$Th nuclear isomer. Such studies are important for the development of a nuclear clock and
for investigations at the interface of atomic and nuclear physics.
\section{Acknowledgments.}
We are grateful to M.~Wiesinger and P.~Thirolf for carefully reading the manuscript and for their valuable comments and suggestions.
This work has been supported in part by NSF
Grant No. PHY-2309254 and the European Research Council (ERC) under the European Union's Horizon 2020 research and innovation program (Grant Agreement No. 856415).
The calculations in this work were done through the use of Information Technologies resources at the University of Delaware, specifically the high-performance Caviness and DARWIN computer clusters.


\begin{thebibliography}{32}%
\makeatletter
\providecommand \@ifxundefined [1]{%
 \@ifx{#1\undefined}
}%
\providecommand \@ifnum [1]{%
 \ifnum #1\expandafter \@firstoftwo
 \else \expandafter \@secondoftwo
 \fi
}%
\providecommand \@ifx [1]{%
 \ifx #1\expandafter \@firstoftwo
 \else \expandafter \@secondoftwo
 \fi
}%
\providecommand \natexlab [1]{#1}%
\providecommand \enquote  [1]{``#1''}%
\providecommand \bibnamefont  [1]{#1}%
\providecommand \bibfnamefont [1]{#1}%
\providecommand \citenamefont [1]{#1}%
\providecommand \href@noop [0]{\@secondoftwo}%
\providecommand \href [0]{\begingroup \@sanitize@url \@href}%
\providecommand \@href[1]{\@@startlink{#1}\@@href}%
\providecommand \@@href[1]{\endgroup#1\@@endlink}%
\providecommand \@sanitize@url [0]{\catcode `\\12\catcode `\$12\catcode
  `\&12\catcode `\#12\catcode `\^12\catcode `\_12\catcode `\%12\relax}%
\providecommand \@@startlink[1]{}%
\providecommand \@@endlink[0]{}%
\providecommand \url  [0]{\begingroup\@sanitize@url \@url }%
\providecommand \@url [1]{\endgroup\@href {#1}{\urlprefix }}%
\providecommand \urlprefix  [0]{URL }%
\providecommand \Eprint [0]{\href }%
\providecommand \doibase [0]{https://doi.org/}%
\providecommand \selectlanguage [0]{\@gobble}%
\providecommand \bibinfo  [0]{\@secondoftwo}%
\providecommand \bibfield  [0]{\@secondoftwo}%
\providecommand \translation [1]{[#1]}%
\providecommand \BibitemOpen [0]{}%
\providecommand \bibitemStop [0]{}%
\providecommand \bibitemNoStop [0]{.\EOS\space}%
\providecommand \EOS [0]{\spacefactor3000\relax}%
\providecommand \BibitemShut  [1]{\csname bibitem#1\endcsname}%
\let\auto@bib@innerbib\@empty
\bibitem [{\citenamefont {Peik}\ and\ \citenamefont {Tamm}(2003)}]{PeiTam03}%
  \BibitemOpen
  \bibfield  {author} {\bibinfo {author} {\bibfnamefont {E.}~\bibnamefont
  {Peik}}\ and\ \bibinfo {author} {\bibfnamefont {C.}~\bibnamefont {Tamm}},\
  }\bibfield  {title} {\bibinfo {title} {Nuclear laser spectroscopy of the 3.5
  e{V} transition in {Th}-229},\ }\href@noop {} {\bibfield  {journal} {\bibinfo
   {journal} {Europhys. Lett.}\ }\textbf {\bibinfo {volume} {61}},\ \bibinfo
  {pages} {181} (\bibinfo {year} {2003})}\BibitemShut {NoStop}%
\bibitem [{\citenamefont {Thielking}\ \emph {et~al.}(2018)\citenamefont
  {Thielking}, \citenamefont {Okhapkin}, \citenamefont {G\l{}owacki},
  \citenamefont {Meier}, \citenamefont {{von der Wense}}, \citenamefont
  {Seiferle}, \citenamefont {D\"{u}llmann}, \citenamefont {Thirolf},\ and\
  \citenamefont {Peik}}]{ThiOkhGlo18}%
  \BibitemOpen
  \bibfield  {author} {\bibinfo {author} {\bibfnamefont {J.}~\bibnamefont
  {Thielking}}, \bibinfo {author} {\bibfnamefont {M.~V.}\ \bibnamefont
  {Okhapkin}}, \bibinfo {author} {\bibfnamefont {P.}~\bibnamefont
  {G\l{}owacki}}, \bibinfo {author} {\bibfnamefont {D.~M.}\ \bibnamefont
  {Meier}}, \bibinfo {author} {\bibfnamefont {L.}~\bibnamefont {{von der
  Wense}}}, \bibinfo {author} {\bibfnamefont {B.}~\bibnamefont {Seiferle}},
  \bibinfo {author} {\bibfnamefont {C.~E.}\ \bibnamefont {D\"{u}llmann}},
  \bibinfo {author} {\bibfnamefont {P.~G.}\ \bibnamefont {Thirolf}},\ and\
  \bibinfo {author} {\bibfnamefont {E.}~\bibnamefont {Peik}},\ }\bibfield
  {title} {\bibinfo {title} {Laser spectroscopic characterization of the
  nuclear-clock isomer $^{229m}${Th}},\ }\href@noop {} {\bibfield  {journal}
  {\bibinfo  {journal} {Nature}\ }\textbf {\bibinfo {volume} {556}},\ \bibinfo
  {pages} {321} (\bibinfo {year} {2018})}\BibitemShut {NoStop}%
\bibitem [{\citenamefont {Tiedau}\ \emph {et~al.}(2024)\citenamefont {Tiedau},
  \citenamefont {Okhapkin}, \citenamefont {Zhang}, \citenamefont {Thielking},
  \citenamefont {Zitzer}, \citenamefont {Peik}, \citenamefont {Schaden},
  \citenamefont {Pronebner}, \citenamefont {Morawetz}, \citenamefont {De~Col},
  \citenamefont {Schneider}, \citenamefont {Leitner}, \citenamefont {Pressler},
  \citenamefont {Kazakov}, \citenamefont {Beeks}, \citenamefont {Sikorsky},\
  and\ \citenamefont {Schumm}}]{TieOkhZha24}%
  \BibitemOpen
  \bibfield  {author} {\bibinfo {author} {\bibfnamefont {J.}~\bibnamefont
  {Tiedau}}, \bibinfo {author} {\bibfnamefont {M.~V.}\ \bibnamefont
  {Okhapkin}}, \bibinfo {author} {\bibfnamefont {K.}~\bibnamefont {Zhang}},
  \bibinfo {author} {\bibfnamefont {J.}~\bibnamefont {Thielking}}, \bibinfo
  {author} {\bibfnamefont {G.}~\bibnamefont {Zitzer}}, \bibinfo {author}
  {\bibfnamefont {E.}~\bibnamefont {Peik}}, \bibinfo {author} {\bibfnamefont
  {F.}~\bibnamefont {Schaden}}, \bibinfo {author} {\bibfnamefont
  {T.}~\bibnamefont {Pronebner}}, \bibinfo {author} {\bibfnamefont
  {I.}~\bibnamefont {Morawetz}}, \bibinfo {author} {\bibfnamefont {L.~T.}\
  \bibnamefont {De~Col}}, \bibinfo {author} {\bibfnamefont {F.}~\bibnamefont
  {Schneider}}, \bibinfo {author} {\bibfnamefont {A.}~\bibnamefont {Leitner}},
  \bibinfo {author} {\bibfnamefont {M.}~\bibnamefont {Pressler}}, \bibinfo
  {author} {\bibfnamefont {G.~A.}\ \bibnamefont {Kazakov}}, \bibinfo {author}
  {\bibfnamefont {K.}~\bibnamefont {Beeks}}, \bibinfo {author} {\bibfnamefont
  {T.}~\bibnamefont {Sikorsky}},\ and\ \bibinfo {author} {\bibfnamefont
  {T.}~\bibnamefont {Schumm}},\ }\bibfield  {title} {\bibinfo {title} {Laser
  excitation of the {Th}-229 nucleus},\ }\href
  {https://doi.org/10.1103/PhysRevLett.132.182501} {\bibfield  {journal}
  {\bibinfo  {journal} {Phys. Rev. Lett.}\ }\textbf {\bibinfo {volume} {132}},\
  \bibinfo {pages} {182501} (\bibinfo {year} {2024})}\BibitemShut {NoStop}%
\bibitem [{\citenamefont {Elwell}\ \emph {et~al.}(2024)\citenamefont {Elwell},
  \citenamefont {Schneider}, \citenamefont {Jeet}, \citenamefont {Terhune},
  \citenamefont {Morgan}, \citenamefont {Alexandrova}, \citenamefont
  {Tran~Tan}, \citenamefont {Derevianko},\ and\ \citenamefont
  {Hudson}}]{ElwSchJee24}%
  \BibitemOpen
  \bibfield  {author} {\bibinfo {author} {\bibfnamefont {R.}~\bibnamefont
  {Elwell}}, \bibinfo {author} {\bibfnamefont {C.}~\bibnamefont {Schneider}},
  \bibinfo {author} {\bibfnamefont {J.}~\bibnamefont {Jeet}}, \bibinfo {author}
  {\bibfnamefont {J.~E.~S.}\ \bibnamefont {Terhune}}, \bibinfo {author}
  {\bibfnamefont {H.~W.~T.}\ \bibnamefont {Morgan}}, \bibinfo {author}
  {\bibfnamefont {A.~N.}\ \bibnamefont {Alexandrova}}, \bibinfo {author}
  {\bibfnamefont {H.~B.}\ \bibnamefont {Tran~Tan}}, \bibinfo {author}
  {\bibfnamefont {A.}~\bibnamefont {Derevianko}},\ and\ \bibinfo {author}
  {\bibfnamefont {E.~R.}\ \bibnamefont {Hudson}},\ }\bibfield  {title}
  {\bibinfo {title} {Laser excitation of the $^{229}\mathrm{Th}$ nuclear
  isomeric transition in a solid-state host},\ }\href@noop {} {\bibfield
  {journal} {\bibinfo  {journal} {Phys. Rev. Lett.}\ }\textbf {\bibinfo
  {volume} {133}},\ \bibinfo {pages} {013201} (\bibinfo {year}
  {2024})}\BibitemShut {NoStop}%
\bibitem [{\citenamefont {Campbell}\ \emph {et~al.}(2011)\citenamefont
  {Campbell}, \citenamefont {Radnaev},\ and\ \citenamefont
  {Kuzmich}}]{CamRadKuz11}%
  \BibitemOpen
  \bibfield  {author} {\bibinfo {author} {\bibfnamefont {C.~J.}\ \bibnamefont
  {Campbell}}, \bibinfo {author} {\bibfnamefont {A.~G.}\ \bibnamefont
  {Radnaev}},\ and\ \bibinfo {author} {\bibfnamefont {A.}~\bibnamefont
  {Kuzmich}},\ }\bibfield  {title} {\bibinfo {title} {Wigner crystals of
  $^{229}\mathrm{Th}$ for optical excitation of the nuclear isomer},\
  }\href@noop {} {\bibfield  {journal} {\bibinfo  {journal} {Phys. Rev. Lett.}\
  }\textbf {\bibinfo {volume} {106}},\ \bibinfo {pages} {223001} (\bibinfo
  {year} {2011})}\BibitemShut {NoStop}%
\bibitem [{\citenamefont {Peik}\ \emph {et~al.}(2021)\citenamefont {Peik},
  \citenamefont {Schumm}, \citenamefont {Safronova}, \citenamefont {P\'alffy},
  \citenamefont {Weitenberg},\ and\ \citenamefont {Thirolf}}]{PeiSchSaf21}%
  \BibitemOpen
  \bibfield  {author} {\bibinfo {author} {\bibfnamefont {E.}~\bibnamefont
  {Peik}}, \bibinfo {author} {\bibfnamefont {T.}~\bibnamefont {Schumm}},
  \bibinfo {author} {\bibfnamefont {M.}~\bibnamefont {Safronova}}, \bibinfo
  {author} {\bibfnamefont {A.}~\bibnamefont {P\'alffy}}, \bibinfo {author}
  {\bibfnamefont {J.}~\bibnamefont {Weitenberg}},\ and\ \bibinfo {author}
  {\bibfnamefont {P.}~\bibnamefont {Thirolf}},\ }\bibfield  {title} {\bibinfo
  {title} {Nuclear clocks for testing fundamental physics},\ }\href@noop {}
  {\bibfield  {journal} {\bibinfo  {journal} {Quantum. Sci. Tech.}\ }\textbf
  {\bibinfo {volume} {6}},\ \bibinfo {pages} {034002} (\bibinfo {year}
  {2021})}\BibitemShut {NoStop}%
\bibitem [{\citenamefont {Kraemer}\ \emph {et~al.}(2023)\citenamefont
  {Kraemer}, \citenamefont {Moens}, \citenamefont {Athanasakis-Kaklamanakis},
  \citenamefont {Bara}, \citenamefont {Beeks}, \citenamefont {Chhetri},
  \citenamefont {Chrysalidis}, \citenamefont {Claessens}, \citenamefont
  {Cocolios}, \citenamefont {Correia}, \citenamefont {Witte}, \citenamefont
  {Ferrer}, \citenamefont {Geldhof}, \citenamefont {Heinke}, \citenamefont
  {Hosseini}, \citenamefont {Huyse}, \citenamefont {K\"oster}, \citenamefont
  {Kudryavtsev}, \citenamefont {Laatiaoui}, \citenamefont {Lica}, \citenamefont
  {Magchiels}, \citenamefont {Manea}, \citenamefont {Merckling}, \citenamefont
  {Pereira}, \citenamefont {Raeder}, \citenamefont {Schumm}, \citenamefont
  {Sels}, \citenamefont {Thirolf}, \citenamefont {Tunhuma}, \citenamefont
  {Bergh}, \citenamefont {Duppen}, \citenamefont {Vantomme}, \citenamefont
  {Verlinde}, \citenamefont {Villarreal}, \citenamefont {Wahl},\ and\
  \citenamefont {Thirolf}}]{KraMoeAtn23}%
  \BibitemOpen
  \bibfield  {author} {\bibinfo {author} {\bibfnamefont {S.}~\bibnamefont
  {Kraemer}}, \bibinfo {author} {\bibfnamefont {J.}~\bibnamefont {Moens}},
  \bibinfo {author} {\bibfnamefont {M.}~\bibnamefont
  {Athanasakis-Kaklamanakis}}, \bibinfo {author} {\bibfnamefont
  {S.}~\bibnamefont {Bara}}, \bibinfo {author} {\bibfnamefont {K.}~\bibnamefont
  {Beeks}}, \bibinfo {author} {\bibfnamefont {P.}~\bibnamefont {Chhetri}},
  \bibinfo {author} {\bibfnamefont {K.}~\bibnamefont {Chrysalidis}}, \bibinfo
  {author} {\bibfnamefont {A.}~\bibnamefont {Claessens}}, \bibinfo {author}
  {\bibfnamefont {T.~E.}\ \bibnamefont {Cocolios}}, \bibinfo {author}
  {\bibfnamefont {J.~G.~M.}\ \bibnamefont {Correia}}, \bibinfo {author}
  {\bibfnamefont {H.~D.}\ \bibnamefont {Witte}}, \bibinfo {author}
  {\bibfnamefont {R.}~\bibnamefont {Ferrer}}, \bibinfo {author} {\bibfnamefont
  {S.}~\bibnamefont {Geldhof}}, \bibinfo {author} {\bibfnamefont
  {R.}~\bibnamefont {Heinke}}, \bibinfo {author} {\bibfnamefont
  {N.}~\bibnamefont {Hosseini}}, \bibinfo {author} {\bibfnamefont
  {M.}~\bibnamefont {Huyse}}, \bibinfo {author} {\bibfnamefont
  {U.}~\bibnamefont {K\"oster}}, \bibinfo {author} {\bibfnamefont
  {Y.}~\bibnamefont {Kudryavtsev}}, \bibinfo {author} {\bibfnamefont
  {M.}~\bibnamefont {Laatiaoui}}, \bibinfo {author} {\bibfnamefont
  {R.}~\bibnamefont {Lica}}, \bibinfo {author} {\bibfnamefont {G.}~\bibnamefont
  {Magchiels}}, \bibinfo {author} {\bibfnamefont {V.}~\bibnamefont {Manea}},
  \bibinfo {author} {\bibfnamefont {C.}~\bibnamefont {Merckling}}, \bibinfo
  {author} {\bibfnamefont {L.~M.~C.}\ \bibnamefont {Pereira}}, \bibinfo
  {author} {\bibfnamefont {S.}~\bibnamefont {Raeder}}, \bibinfo {author}
  {\bibfnamefont {T.}~\bibnamefont {Schumm}}, \bibinfo {author} {\bibfnamefont
  {S.}~\bibnamefont {Sels}}, \bibinfo {author} {\bibfnamefont {P.~G.}\
  \bibnamefont {Thirolf}}, \bibinfo {author} {\bibfnamefont {S.~M.}\
  \bibnamefont {Tunhuma}}, \bibinfo {author} {\bibfnamefont {P.~V.~D.}\
  \bibnamefont {Bergh}}, \bibinfo {author} {\bibfnamefont {P.~V.}\ \bibnamefont
  {Duppen}}, \bibinfo {author} {\bibfnamefont {A.}~\bibnamefont {Vantomme}},
  \bibinfo {author} {\bibfnamefont {M.}~\bibnamefont {Verlinde}}, \bibinfo
  {author} {\bibfnamefont {R.}~\bibnamefont {Villarreal}}, \bibinfo {author}
  {\bibfnamefont {U.}~\bibnamefont {Wahl}},\ and\ \bibinfo {author}
  {\bibfnamefont {P.~G.}\ \bibnamefont {Thirolf}},\ }\bibfield  {title}
  {\bibinfo {title} {Observation of the radiative decay of the $^{229}${Th}
  nuclear clock isomer},\ }\href {https://doi.org/10.1038/s41586-023-05894-z}
  {\bibfield  {journal} {\bibinfo  {journal} {Nature}\ }\textbf {\bibinfo
  {volume} {617}},\ \bibinfo {pages} {706} (\bibinfo {year}
  {2023})}\BibitemShut {NoStop}%
\bibitem [{\citenamefont {Zhang}\ \emph {et~al.}(2024)\citenamefont {Zhang},
  \citenamefont {Ooi}, \citenamefont {Higgins}, \citenamefont {Doyle},
  \citenamefont {von~der Wense}, \citenamefont {Beeks}, \citenamefont
  {Leitner}, \citenamefont {Kazakov}, \citenamefont {Li}, \citenamefont
  {Thirolf}, \citenamefont {Schumm},\ and\ \citenamefont {Ye}}]{ZhaOoiHig24}%
  \BibitemOpen
  \bibfield  {author} {\bibinfo {author} {\bibfnamefont {C.}~\bibnamefont
  {Zhang}}, \bibinfo {author} {\bibfnamefont {T.}~\bibnamefont {Ooi}}, \bibinfo
  {author} {\bibfnamefont {J.~S.}\ \bibnamefont {Higgins}}, \bibinfo {author}
  {\bibfnamefont {J.~F.}\ \bibnamefont {Doyle}}, \bibinfo {author}
  {\bibfnamefont {L.}~\bibnamefont {von~der Wense}}, \bibinfo {author}
  {\bibfnamefont {K.}~\bibnamefont {Beeks}}, \bibinfo {author} {\bibfnamefont
  {A.}~\bibnamefont {Leitner}}, \bibinfo {author} {\bibfnamefont {G.~A.}\
  \bibnamefont {Kazakov}}, \bibinfo {author} {\bibfnamefont {P.}~\bibnamefont
  {Li}}, \bibinfo {author} {\bibfnamefont {P.~G.}\ \bibnamefont {Thirolf}},
  \bibinfo {author} {\bibfnamefont {T.}~\bibnamefont {Schumm}},\ and\ \bibinfo
  {author} {\bibfnamefont {J.}~\bibnamefont {Ye}},\ }\bibfield  {title}
  {\bibinfo {title} {Frequency ratio of the $^{229\mathrm{m}}${Th} nuclear
  isomeric transition and the $^{87}${Sr} atomic clock},\ }\href
  {https://doi.org/10.1038/s41586-024-07839-6} {\bibfield  {journal} {\bibinfo
  {journal} {Nature}\ }\textbf {\bibinfo {volume} {633}},\ \bibinfo {pages}
  {63} (\bibinfo {year} {2024})}\BibitemShut {NoStop}%
\bibitem [{\citenamefont {Huang}\ \emph {et~al.}(2026)\citenamefont {Huang},
  \citenamefont {Yan}, \citenamefont {Xiao}, \citenamefont {Bu}, \citenamefont
  {Zhang}, \citenamefont {Zhao}, \citenamefont {Yan}, \citenamefont {Chen},
  \citenamefont {Zhang}, \citenamefont {Penyazkov}, \citenamefont {Zhan},
  \citenamefont {Yan}, \citenamefont {Wang}, \citenamefont {Li}, \citenamefont
  {Li}, \citenamefont {Qian}, \citenamefont {Liu}, \citenamefont {He},
  \citenamefont {Sun}, \citenamefont {Tian}, \citenamefont {Lu}, \citenamefont
  {Ma}, \citenamefont {Li}, \citenamefont {Wu}, \citenamefont {Gong},
  \citenamefont {Li}, \citenamefont {Shi}, \citenamefont {Li}, \citenamefont
  {Ma}, \citenamefont {Zhu}, \citenamefont {Mo}, \citenamefont {Lin},
  \citenamefont {You}, \citenamefont {Lin}, \citenamefont {Zhang},
  \citenamefont {Hang}, \citenamefont {Su},\ and\ \citenamefont
  {Ding}}]{HuaYanXia26}%
  \BibitemOpen
  \bibfield  {author} {\bibinfo {author} {\bibfnamefont {B.}~\bibnamefont
  {Huang}}, \bibinfo {author} {\bibfnamefont {G.}~\bibnamefont {Yan}}, \bibinfo
  {author} {\bibfnamefont {Q.}~\bibnamefont {Xiao}}, \bibinfo {author}
  {\bibfnamefont {W.}~\bibnamefont {Bu}}, \bibinfo {author} {\bibfnamefont
  {Z.}~\bibnamefont {Zhang}}, \bibinfo {author} {\bibfnamefont
  {C.}~\bibnamefont {Zhao}}, \bibinfo {author} {\bibfnamefont {C.}~\bibnamefont
  {Yan}}, \bibinfo {author} {\bibfnamefont {Z.-A.}\ \bibnamefont {Chen}},
  \bibinfo {author} {\bibfnamefont {P.}~\bibnamefont {Zhang}}, \bibinfo
  {author} {\bibfnamefont {G.}~\bibnamefont {Penyazkov}}, \bibinfo {author}
  {\bibfnamefont {Z.}~\bibnamefont {Zhan}}, \bibinfo {author} {\bibfnamefont
  {L.}~\bibnamefont {Yan}}, \bibinfo {author} {\bibfnamefont {Y.}~\bibnamefont
  {Wang}}, \bibinfo {author} {\bibfnamefont {L.}~\bibnamefont {Li}}, \bibinfo
  {author} {\bibfnamefont {S.}~\bibnamefont {Li}}, \bibinfo {author}
  {\bibfnamefont {X.}~\bibnamefont {Qian}}, \bibinfo {author} {\bibfnamefont
  {X.}~\bibnamefont {Liu}}, \bibinfo {author} {\bibfnamefont {Q.}~\bibnamefont
  {He}}, \bibinfo {author} {\bibfnamefont {T.}~\bibnamefont {Sun}}, \bibinfo
  {author} {\bibfnamefont {H.}~\bibnamefont {Tian}}, \bibinfo {author}
  {\bibfnamefont {B.}~\bibnamefont {Lu}}, \bibinfo {author} {\bibfnamefont
  {N.}~\bibnamefont {Ma}}, \bibinfo {author} {\bibfnamefont {J.}~\bibnamefont
  {Li}}, \bibinfo {author} {\bibfnamefont {Y.}~\bibnamefont {Wu}}, \bibinfo
  {author} {\bibfnamefont {Q.}~\bibnamefont {Gong}}, \bibinfo {author}
  {\bibfnamefont {Y.}~\bibnamefont {Li}}, \bibinfo {author} {\bibfnamefont
  {H.}~\bibnamefont {Shi}}, \bibinfo {author} {\bibfnamefont {X.}~\bibnamefont
  {Li}}, \bibinfo {author} {\bibfnamefont {L.}~\bibnamefont {Ma}}, \bibinfo
  {author} {\bibfnamefont {S.}~\bibnamefont {Zhu}}, \bibinfo {author}
  {\bibfnamefont {Y.}~\bibnamefont {Mo}}, \bibinfo {author} {\bibfnamefont
  {J.}~\bibnamefont {Lin}}, \bibinfo {author} {\bibfnamefont {L.}~\bibnamefont
  {You}}, \bibinfo {author} {\bibfnamefont {Y.}~\bibnamefont {Lin}}, \bibinfo
  {author} {\bibfnamefont {X.}~\bibnamefont {Zhang}}, \bibinfo {author}
  {\bibfnamefont {Y.}~\bibnamefont {Hang}}, \bibinfo {author} {\bibfnamefont
  {L.}~\bibnamefont {Su}},\ and\ \bibinfo {author} {\bibfnamefont
  {S.}~\bibnamefont {Ding}},\ }\href {https://arxiv.org/abs/2606.08870}
  {\bibinfo {title} {A nuclear clock based on $^{229}${Th}}} (\bibinfo {year}
  {2026}),\ \Eprint {https://arxiv.org/abs/2606.08870} {arXiv:2606.08870
  [physics.atom-ph]} \BibitemShut {NoStop}%
\bibitem [{\citenamefont {Col}\ \emph {et~al.}(2026)\citenamefont {Col},
  \citenamefont {Riebner}, \citenamefont {Morawetz}, \citenamefont {Schneider},
  \citenamefont {Sempelmann}, \citenamefont {Schlachet-L\'{e}pinay},
  \citenamefont {Schaden}, \citenamefont {Bartokos}, \citenamefont {Kazakov},
  \citenamefont {Beeks}, \citenamefont {Gerstenecker}, \citenamefont {Pimon},
  \citenamefont {Lahs}, \citenamefont {Hellerschmied}, \citenamefont {Lercher},
  \citenamefont {Premper}, \citenamefont {Niessner}, \citenamefont {Matus},
  \citenamefont {Denker}, \citenamefont {Cizek}, \citenamefont {Cip},
  \citenamefont {Lal}, \citenamefont {Zitzer}, \citenamefont {Petrov},
  \citenamefont {Tiedau}, \citenamefont {Okhapkin}, \citenamefont {Peik},\ and\
  \citenamefont {Schumm}}]{TosRieMor26}%
  \BibitemOpen
  \bibfield  {author} {\bibinfo {author} {\bibfnamefont {L.~T.~D.}\
  \bibnamefont {Col}}, \bibinfo {author} {\bibfnamefont {T.}~\bibnamefont
  {Riebner}}, \bibinfo {author} {\bibfnamefont {I.}~\bibnamefont {Morawetz}},
  \bibinfo {author} {\bibfnamefont {F.}~\bibnamefont {Schneider}}, \bibinfo
  {author} {\bibfnamefont {N.}~\bibnamefont {Sempelmann}}, \bibinfo {author}
  {\bibfnamefont {J.}~\bibnamefont {Schlachet-L\'{e}pinay}}, \bibinfo {author}
  {\bibfnamefont {F.}~\bibnamefont {Schaden}}, \bibinfo {author} {\bibfnamefont
  {M.}~\bibnamefont {Bartokos}}, \bibinfo {author} {\bibfnamefont {G.~A.}\
  \bibnamefont {Kazakov}}, \bibinfo {author} {\bibfnamefont {K.}~\bibnamefont
  {Beeks}}, \bibinfo {author} {\bibfnamefont {B.}~\bibnamefont {Gerstenecker}},
  \bibinfo {author} {\bibfnamefont {M.}~\bibnamefont {Pimon}}, \bibinfo
  {author} {\bibfnamefont {S.}~\bibnamefont {Lahs}}, \bibinfo {author}
  {\bibfnamefont {A.}~\bibnamefont {Hellerschmied}}, \bibinfo {author}
  {\bibfnamefont {T.}~\bibnamefont {Lercher}}, \bibinfo {author} {\bibfnamefont
  {J.}~\bibnamefont {Premper}}, \bibinfo {author} {\bibfnamefont
  {A.}~\bibnamefont {Niessner}}, \bibinfo {author} {\bibfnamefont
  {M.}~\bibnamefont {Matus}}, \bibinfo {author} {\bibfnamefont
  {H.}~\bibnamefont {Denker}}, \bibinfo {author} {\bibfnamefont
  {M.}~\bibnamefont {Cizek}}, \bibinfo {author} {\bibfnamefont
  {O.}~\bibnamefont {Cip}}, \bibinfo {author} {\bibfnamefont {V.}~\bibnamefont
  {Lal}}, \bibinfo {author} {\bibfnamefont {G.}~\bibnamefont {Zitzer}},
  \bibinfo {author} {\bibfnamefont {V.}~\bibnamefont {Petrov}}, \bibinfo
  {author} {\bibfnamefont {J.}~\bibnamefont {Tiedau}}, \bibinfo {author}
  {\bibfnamefont {M.~V.}\ \bibnamefont {Okhapkin}}, \bibinfo {author}
  {\bibfnamefont {E.}~\bibnamefont {Peik}},\ and\ \bibinfo {author}
  {\bibfnamefont {T.}~\bibnamefont {Schumm}},\ }\href
  {https://arxiv.org/abs/2606.04997} {\bibinfo {title} {A thorium-229 optical
  nuclear clock with feedback loop}} (\bibinfo {year} {2026}),\ \Eprint
  {https://arxiv.org/abs/2606.04997} {arXiv:2606.04997 [physics.atom-ph]}
  \BibitemShut {NoStop}%
\bibitem [{\citenamefont {Safronova}\ \emph {et~al.}(2013)\citenamefont
  {Safronova}, \citenamefont {Safronova}, \citenamefont {Radnaev},
  \citenamefont {Campbell},\ and\ \citenamefont {Kuzmich}}]{SafSafRad13}%
  \BibitemOpen
  \bibfield  {author} {\bibinfo {author} {\bibfnamefont {M.~S.}\ \bibnamefont
  {Safronova}}, \bibinfo {author} {\bibfnamefont {U.~I.}\ \bibnamefont
  {Safronova}}, \bibinfo {author} {\bibfnamefont {A.~G.}\ \bibnamefont
  {Radnaev}}, \bibinfo {author} {\bibfnamefont {C.~J.}\ \bibnamefont
  {Campbell}},\ and\ \bibinfo {author} {\bibfnamefont {A.}~\bibnamefont
  {Kuzmich}},\ }\bibfield  {title} {\bibinfo {title} {Magnetic dipole and
  electric quadrupole moments of the $^{229}${Th} nucleus},\ }\href@noop {}
  {\bibfield  {journal} {\bibinfo  {journal} {Phys. Rev. A}\ }\textbf {\bibinfo
  {volume} {88}},\ \bibinfo {pages} {060501(R)} (\bibinfo {year}
  {2013})}\BibitemShut {NoStop}%
\bibitem [{\citenamefont {Porsev}\ \emph {et~al.}(2021)\citenamefont {Porsev},
  \citenamefont {Safronova},\ and\ \citenamefont {Kozlov}}]{PorSafKoz21}%
  \BibitemOpen
  \bibfield  {author} {\bibinfo {author} {\bibfnamefont {S.~G.}\ \bibnamefont
  {Porsev}}, \bibinfo {author} {\bibfnamefont {M.~S.}\ \bibnamefont
  {Safronova}},\ and\ \bibinfo {author} {\bibfnamefont {M.~G.}\ \bibnamefont
  {Kozlov}},\ }\bibfield  {title} {\bibinfo {title} {Precision calculation of
  hyperfine constants for extracting nuclear moments of $^{229}${Th}},\
  }\href@noop {} {\bibfield  {journal} {\bibinfo  {journal} {Phys. Rev. Lett.}\
  }\textbf {\bibinfo {volume} {127}},\ \bibinfo {pages} {253001} (\bibinfo
  {year} {2021})}\BibitemShut {NoStop}%
\bibitem [{\citenamefont {Porsev}\ and\ \citenamefont
  {Safronova}(2021)}]{PorSaf21}%
  \BibitemOpen
  \bibfield  {author} {\bibinfo {author} {\bibfnamefont {S.~G.}\ \bibnamefont
  {Porsev}}\ and\ \bibinfo {author} {\bibfnamefont {M.~S.}\ \bibnamefont
  {Safronova}},\ }\bibfield  {title} {\bibinfo {title} {Role of triple
  excitations in calculating different properties of {Ba}$^+$},\ }\href@noop {}
  {\bibfield  {journal} {\bibinfo  {journal} {Phys. Rev. A}\ }\textbf {\bibinfo
  {volume} {103}},\ \bibinfo {pages} {042815} (\bibinfo {year}
  {2021})}\BibitemShut {NoStop}%
\bibitem [{\citenamefont {Porsev}\ \emph {et~al.}(2020)\citenamefont {Porsev},
  \citenamefont {Safronova}, \citenamefont {Safronova}, \citenamefont
  {Schmidt}, \citenamefont {Bondarev}, \citenamefont {Kozlov}, \citenamefont
  {Tupitsyn},\ and\ \citenamefont {Cheung}}]{PorSafSaf20}%
  \BibitemOpen
  \bibfield  {author} {\bibinfo {author} {\bibfnamefont {S.~G.}\ \bibnamefont
  {Porsev}}, \bibinfo {author} {\bibfnamefont {U.~I.}\ \bibnamefont
  {Safronova}}, \bibinfo {author} {\bibfnamefont {M.~S.}\ \bibnamefont
  {Safronova}}, \bibinfo {author} {\bibfnamefont {P.~O.}\ \bibnamefont
  {Schmidt}}, \bibinfo {author} {\bibfnamefont {A.~I.}\ \bibnamefont
  {Bondarev}}, \bibinfo {author} {\bibfnamefont {M.~G.}\ \bibnamefont
  {Kozlov}}, \bibinfo {author} {\bibfnamefont {I.~I.}\ \bibnamefont
  {Tupitsyn}},\ and\ \bibinfo {author} {\bibfnamefont {C.}~\bibnamefont
  {Cheung}},\ }\bibfield  {title} {\bibinfo {title} {Optical clocks based on
  the {Cf}$^{15+}$ and {Cf}$^{17+}$ ions},\ }\href@noop {} {\bibfield
  {journal} {\bibinfo  {journal} {Phys. Rev. A}\ }\textbf {\bibinfo {volume}
  {102}},\ \bibinfo {pages} {012802} (\bibinfo {year} {2020})}\BibitemShut
  {NoStop}%
\bibitem [{\citenamefont {Porsev}\ and\ \citenamefont
  {Safronova}(2026)}]{PorSaf26}%
  \BibitemOpen
  \bibfield  {author} {\bibinfo {author} {\bibfnamefont {S.~G.}\ \bibnamefont
  {Porsev}}\ and\ \bibinfo {author} {\bibfnamefont {M.~S.}\ \bibnamefont
  {Safronova}},\ }\bibfield  {title} {\bibinfo {title} {Predicting the energies
  of {Cf}$^{17+}$ for an optical clock},\ }\href@noop {} {\bibfield  {journal}
  {\bibinfo  {journal} {Phys. Rev. A}\ }\textbf {\bibinfo {volume} {114}},\
  \bibinfo {pages} {012819} (\bibinfo {year} {2026})}\BibitemShut {NoStop}%
\bibitem [{\citenamefont {Safronova}\ \emph {et~al.}(2018)\citenamefont
  {Safronova}, \citenamefont {Porsev}, \citenamefont {Kozlov}, \citenamefont
  {Thielking}, \citenamefont {Okhapkin}, \citenamefont {G\l{}owacki},
  \citenamefont {Meier},\ and\ \citenamefont {Peik}}]{SafPorKoz18}%
  \BibitemOpen
  \bibfield  {author} {\bibinfo {author} {\bibfnamefont {M.~S.}\ \bibnamefont
  {Safronova}}, \bibinfo {author} {\bibfnamefont {S.~G.}\ \bibnamefont
  {Porsev}}, \bibinfo {author} {\bibfnamefont {M.~G.}\ \bibnamefont {Kozlov}},
  \bibinfo {author} {\bibfnamefont {J.}~\bibnamefont {Thielking}}, \bibinfo
  {author} {\bibfnamefont {M.~V.}\ \bibnamefont {Okhapkin}}, \bibinfo {author}
  {\bibfnamefont {P.}~\bibnamefont {G\l{}owacki}}, \bibinfo {author}
  {\bibfnamefont {D.~M.}\ \bibnamefont {Meier}},\ and\ \bibinfo {author}
  {\bibfnamefont {E.}~\bibnamefont {Peik}},\ }\bibfield  {title} {\bibinfo
  {title} {Nuclear charge radii of $^{229}${Th} from isotope and isomer
  shifts},\ }\href@noop {} {\bibfield  {journal} {\bibinfo  {journal} {Phys.
  Rev. Lett.}\ }\textbf {\bibinfo {volume} {121}},\ \bibinfo {pages} {213001}
  (\bibinfo {year} {2018})}\BibitemShut {NoStop}%
\bibitem [{\citenamefont {Zitzer}\ \emph {et~al.}(2025)\citenamefont {Zitzer},
  \citenamefont {Tiedau}, \citenamefont {D\"ullmann}, \citenamefont
  {Okhapkin},\ and\ \citenamefont {Peik}}]{ZitTieDul25}%
  \BibitemOpen
  \bibfield  {author} {\bibinfo {author} {\bibfnamefont {G.}~\bibnamefont
  {Zitzer}}, \bibinfo {author} {\bibfnamefont {J.}~\bibnamefont {Tiedau}},
  \bibinfo {author} {\bibfnamefont {C.~E.}\ \bibnamefont {D\"ullmann}},
  \bibinfo {author} {\bibfnamefont {M.~V.}\ \bibnamefont {Okhapkin}},\ and\
  \bibinfo {author} {\bibfnamefont {E.}~\bibnamefont {Peik}},\ }\bibfield
  {title} {\bibinfo {title} {Laser spectroscopy on the hyperfine structure and
  isotope shift of sympathetically cooled $^{229}\mathrm{Th}^{3+}$ ions},\
  }\href {https://doi.org/10.1103/PhysRevA.111.L050802} {\bibfield  {journal}
  {\bibinfo  {journal} {Phys. Rev. A}\ }\textbf {\bibinfo {volume} {111}},\
  \bibinfo {pages} {L050802} (\bibinfo {year} {2025})}\BibitemShut {NoStop}%
\bibitem [{\citenamefont {Yamaguchi}\ \emph {et~al.}(2024)\citenamefont
  {Yamaguchi}, \citenamefont {Shigekawa}, \citenamefont {Haba}, \citenamefont
  {Kikunaga}, \citenamefont {Shirasaki}, \citenamefont {Wada},\ and\
  \citenamefont {Katori}}]{YamShiHab24}%
  \BibitemOpen
  \bibfield  {author} {\bibinfo {author} {\bibfnamefont {A.}~\bibnamefont
  {Yamaguchi}}, \bibinfo {author} {\bibfnamefont {Y.}~\bibnamefont
  {Shigekawa}}, \bibinfo {author} {\bibfnamefont {H.}~\bibnamefont {Haba}},
  \bibinfo {author} {\bibfnamefont {H.}~\bibnamefont {Kikunaga}}, \bibinfo
  {author} {\bibfnamefont {K.}~\bibnamefont {Shirasaki}}, \bibinfo {author}
  {\bibfnamefont {M.}~\bibnamefont {Wada}},\ and\ \bibinfo {author}
  {\bibfnamefont {H.}~\bibnamefont {Katori}},\ }\bibfield  {title} {\bibinfo
  {title} {Laser spectroscopy of triply charged $^{229}\mathrm{Th}$ isomer for
  a nuclear clock},\ }\href
  {https://www.nature.com/articles/s41586-024-07296-1} {\bibfield  {journal}
  {\bibinfo  {journal} {Nature}\ }\textbf {\bibinfo {volume} {629}},\ \bibinfo
  {pages} {62} (\bibinfo {year} {2024})}\BibitemShut {NoStop}%
\bibitem [{\citenamefont {Porsev}\ and\ \citenamefont
  {Derevianko}(2006)}]{PorDer06}%
  \BibitemOpen
  \bibfield  {author} {\bibinfo {author} {\bibfnamefont {S.~G.}\ \bibnamefont
  {Porsev}}\ and\ \bibinfo {author} {\bibfnamefont {A.}~\bibnamefont
  {Derevianko}},\ }\bibfield  {title} {\bibinfo {title} {Triple excitations in
  the relativistic coupled-cluster formalism and calculation of {Na}
  properties},\ }\href@noop {} {\bibfield  {journal} {\bibinfo  {journal}
  {Phys. Rev. A}\ }\textbf {\bibinfo {volume} {73}},\ \bibinfo {pages} {012501}
  (\bibinfo {year} {2006})}\BibitemShut {NoStop}%
\bibitem [{\citenamefont {{Dzuba}}\ \emph {et~al.}(1996)\citenamefont
  {{Dzuba}}, \citenamefont {{Flambaum}},\ and\ \citenamefont
  {{Kozlov}}}]{DzuFlaKoz96}%
  \BibitemOpen
  \bibfield  {author} {\bibinfo {author} {\bibfnamefont {V.~A.}\ \bibnamefont
  {{Dzuba}}}, \bibinfo {author} {\bibfnamefont {V.~V.}\ \bibnamefont
  {{Flambaum}}},\ and\ \bibinfo {author} {\bibfnamefont {M.~G.}\ \bibnamefont
  {{Kozlov}}},\ }\bibfield  {title} {\bibinfo {title} {{Combination of the
  many-body perturbation theory with the configuration-interaction method}},\
  }\href@noop {} {\bibfield  {journal} {\bibinfo  {journal} {Phys. Rev. A}\
  }\textbf {\bibinfo {volume} {54}},\ \bibinfo {pages} {3948} (\bibinfo {year}
  {1996})}\BibitemShut {NoStop}%
\bibitem [{\citenamefont {{Safronova}}\ \emph {et~al.}(2009)\citenamefont
  {{Safronova}}, \citenamefont {{Kozlov}}, \citenamefont {{Johnson}},\ and\
  \citenamefont {{Jiang}}}]{SafKozJoh09}%
  \BibitemOpen
  \bibfield  {author} {\bibinfo {author} {\bibfnamefont {M.~S.}\ \bibnamefont
  {{Safronova}}}, \bibinfo {author} {\bibfnamefont {M.~G.}\ \bibnamefont
  {{Kozlov}}}, \bibinfo {author} {\bibfnamefont {W.~R.}\ \bibnamefont
  {{Johnson}}},\ and\ \bibinfo {author} {\bibfnamefont {D.}~\bibnamefont
  {{Jiang}}},\ }\bibfield  {title} {\bibinfo {title} {{Development of a
  configuration-interaction plus all-order method for atomic calculations}},\
  }\href@noop {} {\bibfield  {journal} {\bibinfo  {journal} {\pra}\ }\textbf
  {\bibinfo {volume} {80}},\ \bibinfo {eid} {012516} (\bibinfo {year}
  {2009})}\BibitemShut {NoStop}%
\bibitem [{\citenamefont {Blundell}\ \emph {et~al.}(1989)\citenamefont
  {Blundell}, \citenamefont {Johnson}, \citenamefont {Liu},\ and\ \citenamefont
  {Sapirstein}}]{BluJohLiu89}%
  \BibitemOpen
  \bibfield  {author} {\bibinfo {author} {\bibfnamefont {S.~A.}\ \bibnamefont
  {Blundell}}, \bibinfo {author} {\bibfnamefont {W.~R.}\ \bibnamefont
  {Johnson}}, \bibinfo {author} {\bibfnamefont {Z.~W.}\ \bibnamefont {Liu}},\
  and\ \bibinfo {author} {\bibfnamefont {J.}~\bibnamefont {Sapirstein}},\
  }\bibfield  {title} {\bibinfo {title} {Relativistic all-order calculations of
  energies and matrix elements for {L}i and {B}e$^+$},\ }\href@noop {}
  {\bibfield  {journal} {\bibinfo  {journal} {Phys. Rev. A}\ }\textbf {\bibinfo
  {volume} {40}},\ \bibinfo {pages} {2233} (\bibinfo {year}
  {1989})}\BibitemShut {NoStop}%
\bibitem [{\citenamefont {Blundell}\ \emph {et~al.}(1991)\citenamefont
  {Blundell}, \citenamefont {Johnson},\ and\ \citenamefont
  {Sapirstein}}]{BluJohSap91}%
  \BibitemOpen
  \bibfield  {author} {\bibinfo {author} {\bibfnamefont {S.~A.}\ \bibnamefont
  {Blundell}}, \bibinfo {author} {\bibfnamefont {W.~R.}\ \bibnamefont
  {Johnson}},\ and\ \bibinfo {author} {\bibfnamefont {J.}~\bibnamefont
  {Sapirstein}},\ }\bibfield  {title} {\bibinfo {title} {Relativistic all-order
  calculations of energies and matrix elements in cesium},\ }\href@noop {}
  {\bibfield  {journal} {\bibinfo  {journal} {Phys.\ Rev.\ A}\ }\textbf
  {\bibinfo {volume} {43}},\ \bibinfo {pages} {3407} (\bibinfo {year}
  {1991})}\BibitemShut {NoStop}%
\bibitem [{\citenamefont {Eliav}\ \emph {et~al.}(1996)\citenamefont {Eliav},
  \citenamefont {Kaldor},\ and\ \citenamefont {Ishikawa}}]{EliKalIsh96}%
  \BibitemOpen
  \bibfield  {author} {\bibinfo {author} {\bibfnamefont {E.}~\bibnamefont
  {Eliav}}, \bibinfo {author} {\bibfnamefont {U.}~\bibnamefont {Kaldor}},\ and\
  \bibinfo {author} {\bibfnamefont {Y.}~\bibnamefont {Ishikawa}},\ }\bibfield
  {title} {\bibinfo {title} {Transition energies of barium and radium by the
  relativistic coupled-cluster method},\ }\href@noop {} {\bibfield  {journal}
  {\bibinfo  {journal} {Phys. Rev. A}\ }\textbf {\bibinfo {volume} {53}},\
  \bibinfo {pages} {3050} (\bibinfo {year} {1996})}\BibitemShut {NoStop}%
\bibitem [{\citenamefont {Pal}\ \emph {et~al.}(2007)\citenamefont {Pal},
  \citenamefont {Safronova}, \citenamefont {Johnson}, \citenamefont
  {Derevianko},\ and\ \citenamefont {Porsev}}]{PalSafJoh07}%
  \BibitemOpen
  \bibfield  {author} {\bibinfo {author} {\bibfnamefont {R.}~\bibnamefont
  {Pal}}, \bibinfo {author} {\bibfnamefont {M.~S.}\ \bibnamefont {Safronova}},
  \bibinfo {author} {\bibfnamefont {W.~R.}\ \bibnamefont {Johnson}}, \bibinfo
  {author} {\bibfnamefont {A.}~\bibnamefont {Derevianko}},\ and\ \bibinfo
  {author} {\bibfnamefont {S.~G.}\ \bibnamefont {Porsev}},\ }\bibfield  {title}
  {\bibinfo {title} {Relativistic coupled-cluster single-double method applied
  to alkali-metal atoms},\ }\href@noop {} {\bibfield  {journal} {\bibinfo
  {journal} {Phys. Rev. A}\ }\textbf {\bibinfo {volume} {75}},\ \bibinfo
  {pages} {042515} (\bibinfo {year} {2007})}\BibitemShut {NoStop}%
\bibitem [{\citenamefont {Sahoo}\ \emph {et~al.}(2007)\citenamefont {Sahoo},
  \citenamefont {Das}, \citenamefont {Chaudhuri},\ and\ \citenamefont
  {Mukherjee}}]{SahDasCha07}%
  \BibitemOpen
  \bibfield  {author} {\bibinfo {author} {\bibfnamefont {B.~K.}\ \bibnamefont
  {Sahoo}}, \bibinfo {author} {\bibfnamefont {B.~P.}\ \bibnamefont {Das}},
  \bibinfo {author} {\bibfnamefont {R.~K.}\ \bibnamefont {Chaudhuri}},\ and\
  \bibinfo {author} {\bibfnamefont {D.}~\bibnamefont {Mukherjee}},\ }\bibfield
  {title} {\bibinfo {title} {Theoretical studies of the $6s\,\,^2\!{S}_{1/2}
  \rightarrow 5d\,\,^2\!{D}_{3/2}$ parity-nonconserving transition amplitude in
  {B}a$^+$ and associated properties},\ }\href@noop {} {\bibfield  {journal}
  {\bibinfo  {journal} {Phys. Rev. A}\ }\textbf {\bibinfo {volume} {75}},\
  \bibinfo {pages} {032507} (\bibinfo {year} {2007})}\BibitemShut {NoStop}%
\bibitem [{\citenamefont {Safronova}\ \emph {et~al.}(2014)\citenamefont
  {Safronova}, \citenamefont {Dzuba}, \citenamefont {Flambaum}, \citenamefont
  {Safronova}, \citenamefont {Porsev},\ and\ \citenamefont
  {Kozlov}}]{SafDzuFla14PRA1}%
  \BibitemOpen
  \bibfield  {author} {\bibinfo {author} {\bibfnamefont {M.~S.}\ \bibnamefont
  {Safronova}}, \bibinfo {author} {\bibfnamefont {V.~A.}\ \bibnamefont
  {Dzuba}}, \bibinfo {author} {\bibfnamefont {V.~V.}\ \bibnamefont {Flambaum}},
  \bibinfo {author} {\bibfnamefont {U.~I.}\ \bibnamefont {Safronova}}, \bibinfo
  {author} {\bibfnamefont {S.~G.}\ \bibnamefont {Porsev}},\ and\ \bibinfo
  {author} {\bibfnamefont {M.~G.}\ \bibnamefont {Kozlov}},\ }\bibfield  {title}
  {\bibinfo {title} {Highly charged {Ag}-like and {In}-like ions for the
  development of atomic clocks and the search for $\alpha$ variation},\
  }\href@noop {} {\bibfield  {journal} {\bibinfo  {journal} {Phys. Rev. A}\
  }\textbf {\bibinfo {volume} {90}},\ \bibinfo {pages} {042513} (\bibinfo
  {year} {2014})}\BibitemShut {NoStop}%
\bibitem [{\citenamefont {Tupitsyn}\ \emph {et~al.}(2016)\citenamefont
  {Tupitsyn}, \citenamefont {Kozlov}, \citenamefont {Safronova}, \citenamefont
  {Shabaev},\ and\ \citenamefont {Dzuba}}]{TupKozSaf16}%
  \BibitemOpen
  \bibfield  {author} {\bibinfo {author} {\bibfnamefont {I.~I.}\ \bibnamefont
  {Tupitsyn}}, \bibinfo {author} {\bibfnamefont {M.~G.}\ \bibnamefont
  {Kozlov}}, \bibinfo {author} {\bibfnamefont {M.~S.}\ \bibnamefont
  {Safronova}}, \bibinfo {author} {\bibfnamefont {V.~M.}\ \bibnamefont
  {Shabaev}},\ and\ \bibinfo {author} {\bibfnamefont {V.~A.}\ \bibnamefont
  {Dzuba}},\ }\bibfield  {title} {\bibinfo {title} {Quantum electrodynamical
  shifts in multivalent heavy ions},\ }\href@noop {} {\bibfield  {journal}
  {\bibinfo  {journal} {Phys. Rev. Lett.}\ }\textbf {\bibinfo {volume} {117}},\
  \bibinfo {pages} {253001} (\bibinfo {year} {2016})}\BibitemShut {NoStop}%
\bibitem [{Ral()}]{RalKraRea11}%
  \BibitemOpen
  \href@noop {} {}\bibinfo {note} {Yu.~Ralchenko, A.~Kramida, J.~Reader, and
  the NIST ASD Team (2011). NIST Atomic Spectra Database (version 4.1).
  Available at http://physics.nist.gov/asd. National Institute of Standards and
  Technology, Gaithersburg, MD.}\BibitemShut {Stop}%
\bibitem [{ThI()}]{ThIV}%
  \BibitemOpen
  \href@noop {} {}\bibinfo {howpublished}
  {\url{http://www.lac.universite-paris-saclay.fr/Data/Database/Tab-energy/Thorium/Th-el-dir.html}}\BibitemShut
  {NoStop}%
\bibitem [{\citenamefont {Li}\ \emph {et~al.}(2021)\citenamefont {Li},
  \citenamefont {Qiao}, \citenamefont {Tang},\ and\ \citenamefont
  {Shi}}]{LiQiaTan21}%
  \BibitemOpen
  \bibfield  {author} {\bibinfo {author} {\bibfnamefont {F.-C.}\ \bibnamefont
  {Li}}, \bibinfo {author} {\bibfnamefont {H.-X.}\ \bibnamefont {Qiao}},
  \bibinfo {author} {\bibfnamefont {Y.-B.}\ \bibnamefont {Tang}},\ and\
  \bibinfo {author} {\bibfnamefont {T.-Y.}\ \bibnamefont {Shi}},\ }\bibfield
  {title} {\bibinfo {title} {Relativistic coupled-cluster calculation of
  hyperfine-structure constants of $^{229}\mathrm{Th}^{3+}$ and evaluation of
  the electromagnetic nuclear moments of $^{229}\mathrm{Th}$},\ }\href@noop {}
  {\bibfield  {journal} {\bibinfo  {journal} {Phys. Rev. A}\ }\textbf {\bibinfo
  {volume} {104}},\ \bibinfo {pages} {062808} (\bibinfo {year}
  {2021})}\BibitemShut {NoStop}%
\bibitem [{\citenamefont {Korol}\ and\ \citenamefont
  {Kozlov}(2007)}]{KorKoz07}%
  \BibitemOpen
  \bibfield  {author} {\bibinfo {author} {\bibfnamefont {V.~A.}\ \bibnamefont
  {Korol}}\ and\ \bibinfo {author} {\bibfnamefont {M.~G.}\ \bibnamefont
  {Kozlov}},\ }\bibfield  {title} {\bibinfo {title} {Relativistic corrections
  to the isotope shift in light ions},\ }\href@noop {} {\bibfield  {journal}
  {\bibinfo  {journal} {Phys. Rev. A}\ }\textbf {\bibinfo {volume} {76}},\
  \bibinfo {pages} {022103} (\bibinfo {year} {2007})}\BibitemShut {NoStop}%
\end{thebibliography}
%

\end{document}